\documentclass[runningheads]{llncs}

\usepackage[T1]{fontenc}
\usepackage{tabularx}
\usepackage{array}
\usepackage{graphicx}
\usepackage{amsmath,amssymb,amsfonts}
\usepackage{algorithmic}
\usepackage{textcomp}
\usepackage{xcolor}
\usepackage{pdflscape}
\usepackage{longtable}
\usepackage{booktabs}
\usepackage{algorithm}
\usepackage{multirow}
\usepackage{placeins}
\usepackage{pgfplots}
\usepackage[colorlinks=true, linkcolor=blue]{hyperref}
\begin{document}
\title{High-Efficiency Split Computing for Cooperative Edge Systems: A Novel Compressed Sensing Bottleneck}
\titlerunning{High-Efficiency Compressed Sensing Bottleneck}
%
%
\author{Hailin Zhong \and Donglong Chen\inst{*}}
\authorrunning{H. Zhong and D. Chen}
%
\institute{Beijing Normal-Hong Kong Baptist University, Zhuhai, Guangdong Province, China
\email{r130026215@mail.uic.edu.cn, donglongchen@uic.edu.cn}
}
\maketitle              
\begin{abstract}
The advent of big data and AI has precipitated a demand for computational frameworks that ensure real-time performance, accuracy, and privacy. While edge computing mitigates latency and privacy concerns, its scalability is constrained by the resources of edge devices, thus prompting the adoption of split computing (SC) addresses these limitations. However, SC faces challenges in (1) efficient data transmission under bandwidth constraints and (2) balancing accuracy with real-time performance. To tackle these challenges, we propose a novel split computing architecture inspired by compressed sensing (CS) theory. At its core is the \textit{High-Efficiency Compressed Sensing Bottleneck (HECS-B)}, which incorporates an efficient compressed sensing autoencoder into the shallow layer of a deep neural network (DNN) to create a bottleneck layer using the knowledge distillation method. This bottleneck splits the DNN into a distributed model while efficiently compressing intermediate feature data, preserving critical information for seamless reconstruction in the cloud.

Through rigorous theoretical analysis and extensive experimental validation in both simulated and real-world settings, we demonstrate the effectiveness of the proposed approach. Compared to state-of-the-art methods, our architecture reduces bandwidth utilization by \textbf{50\%}, maintains high accuracy, and achieves a \textbf{60\% speed-up} in computational efficiency. The results highlight significant improvements in bandwidth efficiency, processing speed, and model accuracy, underscoring the potential of HECS-B to bridge the gap between resource-constrained edge devices and computationally intensive cloud services. 

\end{abstract}
\section{Introduction}
With the rapid advancements in big data and artificial intelligence (AI), coupled with the proliferation of interconnected devices and the growing demand for intelligent applications, real-time performance, computational efficiency, and privacy preservation have become critical challenges~\cite{furtuanpey2024frankensplit}. Traditional local computing~\cite{matsubara2022towards} is constrained by the limited computational capacity of edge devices, while cloud computing, though capable of providing immense computational power, suffers from inherent high latency, data transmission overhead, and potential privacy risks. This dichotomy underscores the need for a paradigm that can balance computational power, low latency, and privacy requirements, making \textit{edge computing} a promising research direction~\cite{matsubara2022towards}. By bringing computation closer to the data source, edge computing demonstrates significant potential in mitigating latency and safeguarding user privacy~\cite{wang2020}.

Nevertheless, the scalability of edge computing is hindered by the finite computational resources available at edge devices. To address this, \textit{split computing (SC)}~\cite{matsubara2022towards} has emerged as a viable solution, leveraging a collaborative division of computational tasks between edge devices and the cloud to optimize resource utilization. However, split computing faces two primary challenges~\cite{matsubara2022towards}: (1) \textit{efficient data transmission under bandwidth constraints} and (2) \textit{maintaining high accuracy while ensuring real-time performance}. Solving these challenges is crucial to unlocking the full potential of split computing in practical applications.

Recent studies have made substantial progress in addressing these challenges. For instance, Alireza Furutanpey {et al.} proposed the \textit{Shallow Variational Bottleneck (SVB)}~\cite{furtuanpey2024frankensplit} architecture, which integrates a variational autoencoder to enhance SC performance. Their approach~\cite{furtuanpey2024frankensplit} achieved a {60\% reduction in bandwidth utilization} compared to state-of-the-art SC methods while improving processing speed by a factor of {16}. Despite these advancements, existing methods encounter limitations when applied to large-scale datasets or scenarios demanding ultra-low latency, highlighting the need for further innovations.

Motivated by these challenges, this work draws inspiration from \textit{compressed sensing (CS)} theory in signal processing and introduces a novel split computing architecture. At its core, the proposed architecture incorporates a \textit{High-Efficiency Compressed Sensing Bottleneck (HECS-B)}, which introduces a high-efficiency compressed sensing autoencoder and inserts it into the shallow layer of DNN to form a bottleneck layer by knowledge distillation method, as a pivotal component for model reconstruction. 

We open source our code of this work in \href{https://github.com/}{GitHub} for future reproduction and extension. The key contributions of this work can be summarized as follows:
\begin{enumerate}
    \item Initially, we applied compressed sensing theory to the field of edge computing and innovatively designed a high-efficiency compressed sensing autoencoder to serve as the bottleneck layer of the split model.
    \item A novel split computing architecture by combining \textit{Compressed Sensing Theory, Splitting Computing Theory}, and \textit{Knowledge Distillation Theory}, introducing the \textit{High-Efficiency Compressed Sensing Bottleneck (HECS-B)}, is proposed. The HECS-B uses an encoder to map the intermediate features of the DNN to the latent space for compression.  The compressed features are then transmitted to a decoder for reconstruction, optimizing both bandwidth utilization and computational performance.
    \item We deploy the architecture in real-world scenarios, with experimental results showcasing significant breakthroughs:
    \begin{itemize}
    \item Substantially reduced bandwidth requirements, with a {50\% reduction} compared to SVB methods.
    \item Maintained model accuracy, with no degradation in performance.
    \item Enhanced computational efficiency, achieving {60\% speed-up}, drastically reducing system latency.
    \end{itemize}
    
\end{enumerate}

The remaining of this paper is organized as follows: {Section 2} reviews related work in split computing and model partitioning. {Section 3} introduces the theoretical foundations underlying the proposed architecture and details the design, implementation, and optimization of the proposed split computing framework. {Section 4} provides a comprehensive evaluation, including mathematical analysis, experimental results validating the effectiveness of HECS-B, and deployment in real-world scenarios, highlighting the superiority of our method. {Section 5} concludes the paper and outlines future research directions.

\section{Related Work}

\subsection{Edge Computing}
\begin{figure}
  \centering
  \includegraphics[width=1\textwidth]{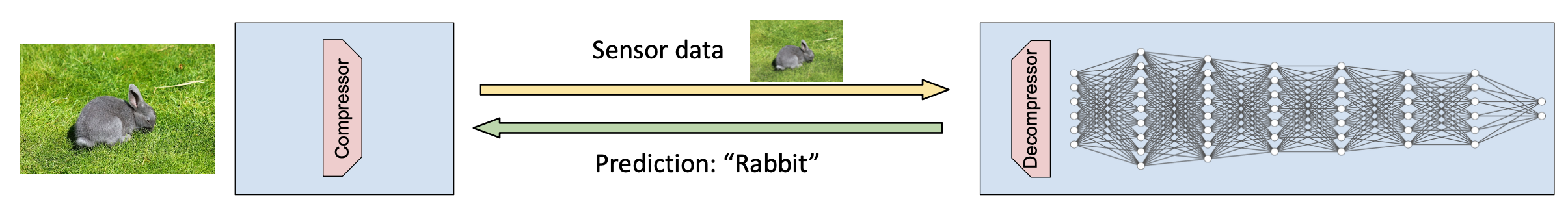}
  \caption{Edge computing structure.}
  \label{fig:f2}
\end{figure}

Edge computing offloads computational responsibilities from the local to the server. In Figure \ref{fig:f2}, which has show how edge computing work, the data be collected and compressed in local device and sends it to the edge server, after the model inference on the server, the inference results will be send back to the local device. However, transmitting the input \( x \) in its entirety poses significant challenges, particularly in scenarios with unstable network conditions, leading to potential delays or task failures. Compressing input data with formats such as JPEG can reduce transmission times, but these formats are primarily designed for signal reconstruction, which might expose sensitive data and raise privacy concerns~\cite{wang2020}. Furthermore, these formats generally consume more bandwidth than task-optimized compressed representations, such as those implemented in bottleneck-based split computing frameworks, which will be elaborated on later.

\subsection{Split Computing}

\begin{figure}
    \centering
    \includegraphics[width=1\textwidth]{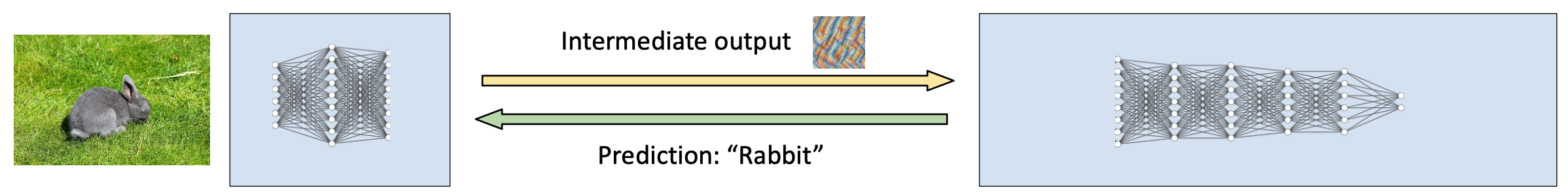}
    \caption{Split computing structure}
    \label{fig:f3}
\end{figure}
In Figure~\ref{fig:f3} has show the how split computing work, the model be split into two parts, while the head part put in local device and the tail part put in the server device, then these two devices cooperatively inference to finish the task. Split computing have two key goals~\cite{matsubara2019}: (i) allocating computational tasks between edge and servers, (ii) minimize transmission delays. For a neural network \( M(\cdot) \) comprising \( L \) layers, the intermediate output at the \( \ell \)-th layer is represented as \( z_\ell \). Early approaches to split computing partition \( M(\cdot) \) by selecting a specific layer \( \ell \), dividing the computation into two segments: \( z_\ell = M_H(x) \), processed on the mobile device, and \( \hat{y} = M_T(z_\ell) \), executed at the edge server.

\subsubsection{Without Bottleneck Injection}
\begin{figure}
    \centering
    \includegraphics[width=0.7\textwidth]{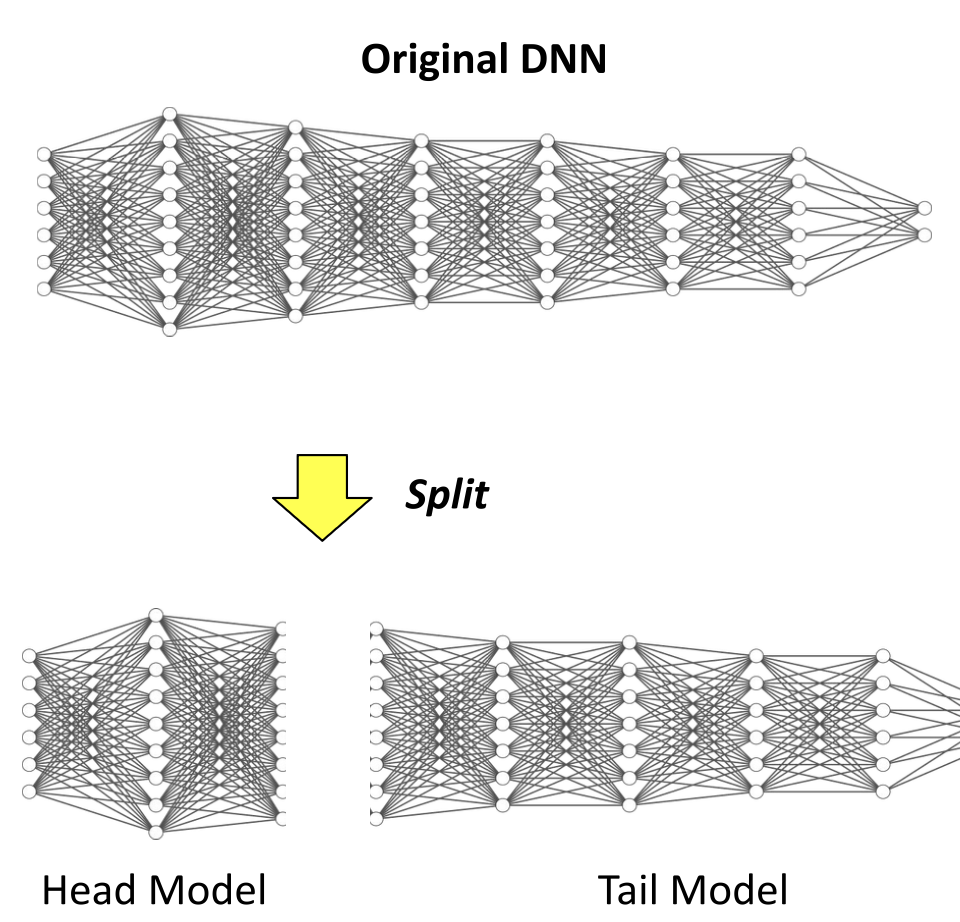}
    \caption{Split computing without bottleneck injection}
    \label{fig:f4}
\end{figure}
Early approaches~\cite{matsubara2019} to split computing directly allocate the head and tail submodels respectively. In Figure~\ref{fig:f4} has show the model structure which be splited directly(without bottleneck inject). While this straightforward design maintains the original network's accuracy, it delegates a portion of the computational workload to the mobile device, which frequently lacks sufficient processing capacity compared to the server. As a result, it may extended overall execution times. The transmission time for \( z_\ell \), relative to the input \( x \), is influenced by the size of \( z_\ell \). However, in most real-world use cases, \( z_\ell \) only shows substantial size reduction at deeper layers of the network, thereby increasing the computational burden on the mobile device.

\subsubsection{With Bottleneck Injection}
\begin{figure}
    \centering
    \includegraphics[width=0.7\textwidth]{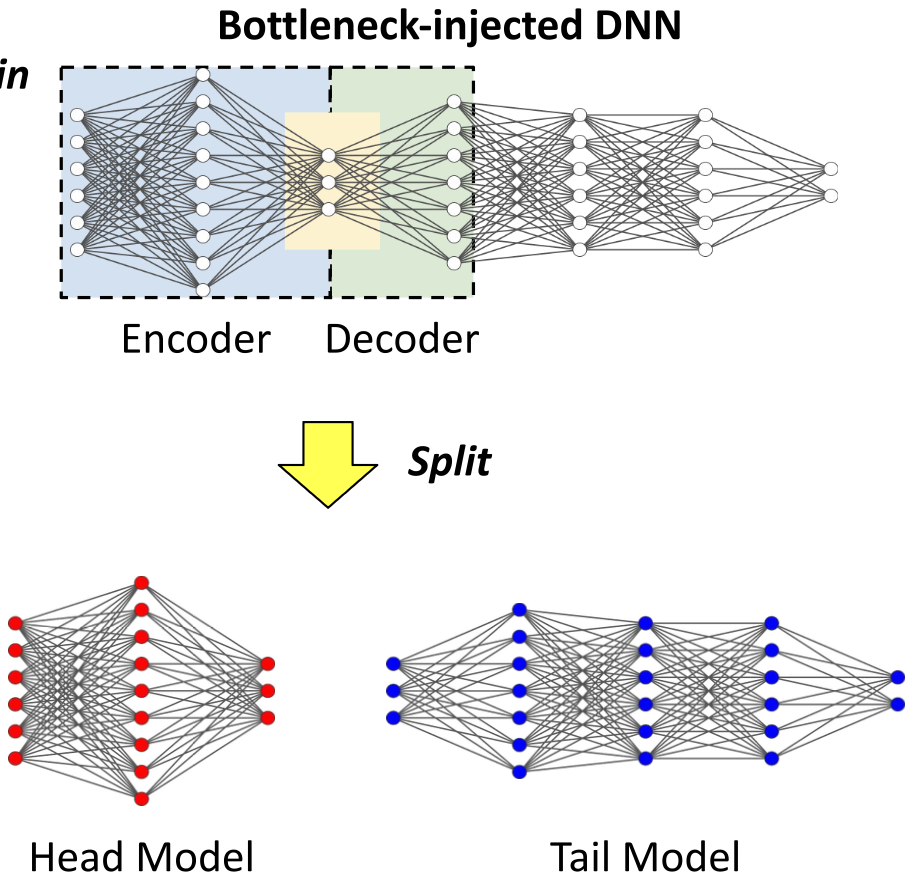}
    \caption{Split computing with bottleneck injection}
    \label{fig:f5}
\end{figure}
Recent advancements~\cite{matsubara2019} in split computing often employ bottleneck layers to implement compression strategies designed for specific applications~\cite{matsubara2019}. This method segments the model into three distinct components: \( M_E \), \( M_D \), and \( M_T \). In Figure~\ref{fig:f5} has show the model structure which be spitted with bottleneck inject. In detail, for a given input \( x \), the intermediate output produced at \( \ell \)-th layer of original model which is denoted as \( z_\ell|x \). The initial submodel \( M_E \) computes this intermediate result, which is subsequently compressed by \( M_D \) into a smaller representation \( \hat{z}_\ell|x \). This compressed data is transmitted to the edge server, where \( M_T \) processes it to generate the final prediction \( \hat{y} \). System performance is assessed by comparing the accuracy of the generated output to the ground truth \( y \). In this setup, \( M_E \) operates on the mobile device, while \( M_D \) and \( M_T \) are deployed on the edge server. The communication channel carries the compressed tensor \( \hat{z}_\ell|x \), with the bottleneck layer positioned between \( M_E \) and \( M_D \), serving as a critical component for balancing data compression and prediction accuracy.

While the inclusion of a bottleneck layer enhances split computing by reducing the size of transmitted intermediate data, it introduces a trade-off between transmission efficiency and model accuracy. Optimizing this balance requires careful adjustment of the compression process to meet the specific demands of the target application or task.

\subsection{Model Compression}

Model compression plays a vital role in deep learning, aiming to minimize the computational complexity and storage requirements of models without significantly compromising their performance. Techniques such as quantization, pruning, and knowledge distillation have emerged as highly effective solutions and are widely used in practical scenarios for their ability to balance efficiency and accuracy.

\subsubsection{Quantization and Pruning}

Quantization and pruning~\cite{han2015learning,han2016deep,jacob2018quantization,li2020train} are foundational techniques in model compression. Quantization focuses on reducing the numerical precision of model parameters, typically by lowering the bit-width, to enhance storage efficiency and computational speed. Pruning, on the other hand, aims to streamline the model by eliminating redundant parameters, thereby simplifying its structure. Unlike strategies that involve directly designing compact models, these methods usually begin with training a larger, more complex model, which is then compressed for deployment. Jacob et al.~\cite{jacob2018quantization} have shown that their quantization method significantly enhances the balance between inference speed and accuracy in MobileNet~\cite{howard2017mobilenets}, outperforming float-only MobileNet on Snapdragon 835 and 821 processors.

Despite its advantages, pruning presents notable challenges when implemented on general-purpose hardware. Research by Li et al.~\cite{li2017infogail} and Liu et al.~\cite{liu2021ebert} points out that the irregular sparsity introduced during pruning often complicates both inference acceleration and hardware optimization. Consequently, while pruning effectively reduces the model's memory footprint, its practical impact on inference efficiency may be constrained.

\subsubsection{Knowledge Distillation}
\begin{figure}
    \centering
    \includegraphics[width=0.7\textwidth]{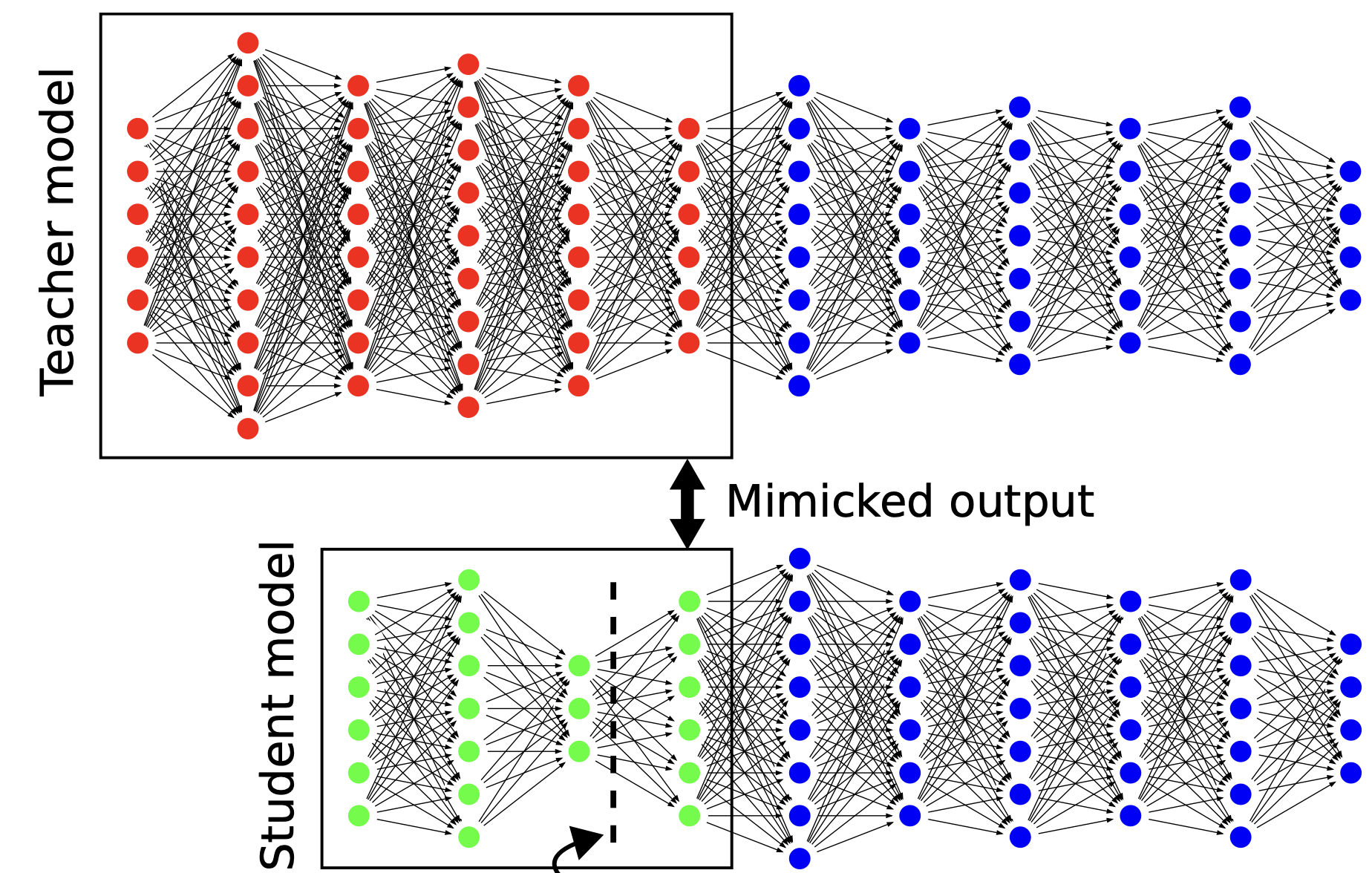}
    \caption{Original(teacher) model and target(student) model structure of KD}
    \label{fig:f6}
\end{figure}
In Figure~\ref{fig:f6}, which has show the how knowledge distillation (KD) method work, for a original (teacher) model, we design a compressed (student) model, then  use the teacher's output as soft label of the training data to train the student together. Which provides the unique way for compressing models from a pre-trained "teacher" model to a "student" model. Rather than directly reducing the parameter count, this approach leverages the predictions of the teacher model, use it to guide the student model learning process. As results, the student model achieves a level of accuracy comparable to that of the teacher while maintaining a simpler and more efficient architecture. This makes knowledge distillation particularly suitable for enhancing lightweight models, as it improves their accuracy.

Ba and Caruana~\cite{ba2014deep} propose techniques where smaller networks learn to mimic the output of a larger model. Their results shows student models trained using KD method can match the performance of deeper networks in tasks such as phoneme recognition and image classification. These findings underscore the ability of knowledge distillation to effectively balance model simplicity with predictive performance.

\subsection{Feature Compression}

Feature compression is essential in split computing, as it reduces the size of intermediate feature representations to facilitate efficient data transmission. One widely used method for feature compression is the encoder-decoder framework. In this configuration, the encoder operates on the mobile device to compress features by decreasing their dimensionality. These compressed features are subsequently sent to the edge server, where the decoder reconstructs them for further computational processing.

\subsubsection{Autoencoder-Based Compression}
\begin{figure}
    \centering
    \includegraphics[width=1\textwidth]{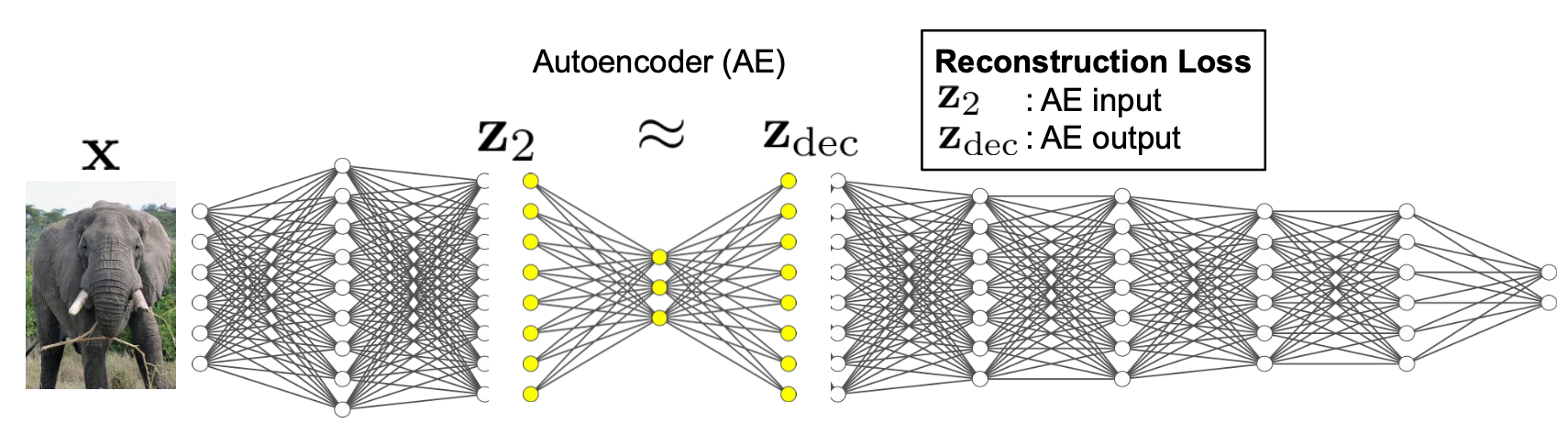}
    \caption{Autoencoder based feature compression}
    \label{fig:f7}
\end{figure}
Autoencoders are extensively used for feature compression in split computing. In the Figure~\ref{fig:f7} has show the structure of the split model with autoencoder used for mid-feature compression. In a instance, Variational autoencoders (VAEs)~\cite{furtuanpey2024frankensplit} map features into a probabilistic latent space, enabling the creation of compact and informative representations. Additionally, denoising autoencoders improve robustness against transmission noise, which is particularly beneficial in wireless edge environments~\cite{balle2017end}.

While significant progress has been made in edge computing, split computing, and feature compression, achieving an optimal balance between communication efficiency and model accuracy remains an open challenge in split computing. Existing methods, including various feature compression techniques and bottleneck designs, have laid a solid foundation, but the integration of compressed sensing principles into split computing frameworks has not been adequately explored. 

To address this gap, this study introduces, for the first time, a novel split computing framework leveraging compressed sensing. Specifically, we propose an High-Efficiency Compressed Sensing Bottleneck that maximizes compression rate and transmission efficiency while maintaining model accuracy. The following methodology section elaborates on the theoretical foundations, design principles, and implementation strategies of this framework.

\section{Methodology}

While we discussing the specifics of the proposed architecture, we present a concise overview of the theoretical foundations that demonstrate its practicality. In this work, probability distributions are represented using uppercase letters, with their corresponding absolutely continuous densities (defined with respect to an appropriate reference measure) denoted by lowercase letters. Similarly, we use uppercase notation for random variables and lowercase notation for their specific realizations.

\subsection{High-Efficiency Compressed Sensing Autoencoder (HECSA) Design}
\subsubsection{Compressed Sensing (CS)}

Compressed sensing~\cite{1614066} is a technique for reconstructing a high-dimensional data point \( X \in \mathbb{R}^n \) using a limited set of measurements \( Y \in \mathbb{R}^m \), where \( m < n \). This implies that fewer measurements are utilized than the original dimensionality of the data. The relationship between \( X \) and \( Y \) is established via the matrix \( W \in \mathbb{R}^{m \times l} \) and the function \( f_\psi : \mathbb{R}^n \to \mathbb{R}^l \) (an integer is \( l > 0 \) ), which combine to form the following equation:
\begin{equation}
    y = W f_\psi(x) + \epsilon,
\end{equation}
the \( \epsilon \) represents noise in measurements.

If the acquisition function \( f_\psi(\cdot) \) is set as the identity mapping (i.e., \( f_\psi(x) = x \)), the problem simplifies to solving underdetermined linear systems, where \( y \) represents a noisy linear projection of \( x \). In more complex scenarios, \( f_\psi(\cdot) \) can be tailored to transform \( x \) into a representation \( f_\psi(x) \) that is optimized for compressed sensing. For example, \( f_\psi(\cdot) \) might perform a basis transformation, such as projecting onto the Fourier basis to exploit signal sparsity in audio data. Notably, the output dimensionality of \( f_\psi(\cdot) \) (codomain) does not have to match the input space, accommodating cases where \( l \neq n \).

\subsubsection{Fundamental Assumptions based on CS}

To describe the relationship between signals \( X \) and their corresponding measurements \( Y \), we define a joint probability distribution \( Q_\phi(X, Y) \), which can be factorized by:
\begin{equation}
    Q_\phi(X, Y) = Q_{\text{data}}(X) Q_\phi(Y | X),
\end{equation}
The \( Q_{\text{data}}(X) \) represents distribution of the data, \( Q_\phi(Y | X) \) denotes a conditional incorporates measure\( \epsilon \). The parameter set \( \phi \) encompasses the parameters of both \( W \) and \( \psi \).

As an example, when the measurement noise \( \epsilon \) follows an isotropic Gaussian distribution with constant variance \( \sigma^2 \), \( Q_\phi(Y | X) \) expressed by:
\begin{equation}
    Q_\phi(Y | X) = \mathcal{N}(W f_\psi(X), \sigma^2 I_m),
\end{equation}
where \( \mathcal{N} \) denotes the multivariate Gaussian distribution, \( W f_\psi(X) \) specifies the mean vector, and \( \sigma^2 I_m \) defines the covariance matrix.

This joint probability expression serves as the foundation for information-theoretic analysis in task-agnostic compressed sensing. In recent works~\cite{Baraniuk2008CS}, maximizing the mutual information between the original signal \( X \) and its compressed representation \( Y \) has been shown to yield representations that retain informative content necessary for various downstream tasks, even in the absence of task supervision.

Based on the theory of CS, then we combine it with autoencoder to improve it, and thus design a new HECSA.

\subsubsection{Autoencoder}

An autoencoder~\cite{ng2011sparse} consists of two parameterized functions \( (e, d) \), which work in tandem to encode and decode data points. The encoder \( e : \mathbb{R}^n \to \mathbb{R}^m \) maps an input from an \( n \)-dimensional space to a compressed representation within an \( m \)-dimensional latent space. Conversely, the decoder \( d : \mathbb{R}^m \to \mathbb{R}^n \) reconstructs the original input data from this latent representation.

The training process of autoencoder aims for minimize reconstruction error with datasets \( D \), ensuring that the reconstructed output closely approximates the original input. This optimization can be formulated by:
\begin{equation}
    \min_{e, d} \sum_{x \in D} \|x - d(e(x))\|_2^2,
\end{equation}
Both \( e(\cdot) \) and \( d(\cdot) \) are commonly implemented as neural networks. Through this optimization, the autoencoder learns a compact, low-dimensional representation that preserves key features of the input data while minimizing reconstruction loss.

\subsubsection{HECSA Optimization Objective}

The High-Efficiency Compressed Sensing Bottleneck framework aims to optimize the parameters \( \phi \) to facilitate accurate and efficient signal recovery of \( X \) from its corresponding measurements \( Y \). The core objective is to max transfer from \( X \) and \( Y \), which is formulated by:
\begin{equation}
    \max_\phi I_\phi(X, Y) = \int q_\phi(x, y) \log \frac{q_\phi(x, y)}{q_{\text{data}}(x) q_\phi(y)} \, dx \, dy,
\end{equation}
The \( I_\phi(X, Y) \) denotes the mutual information, the \( q_\phi(x, y) \) is statistic distribution of the \( X \) and the \( Y \). Alternatively, this mutual information can be described using differential entropy:
\begin{equation}
    I_\phi(X, Y) = H(X) - H_\phi(X | Y),
\end{equation}
\( H \) is entropy which of data distribution. The objective is to retain as much information as possible in \( Y \) about \( X \), thereby minimizing reconstruction error during the recovery process.

This optimization can also be viewed as max the expected log posterior probability \( X \) to \( Y \). Since the entropy \( H(X) \) is independent of \( \phi \), it can be excluded from the optimization, resulting:
\begin{equation}
    \max_\phi -H_\phi(X | Y) = \mathbb{E}_{Q_\phi(X, Y)}[\log q_\phi(x | y)],
\end{equation}
where \( -H_\phi(X | Y) \) quantifies the conditional-entropy \( X \) to \( Y \). This reformulation emphasizes the goal of learning parameters \( \phi \) that maximize the posterior probability of \( X \), thus enhancing reconstruction accuracy.

It is important to note that maximizing mutual information does not necessarily equate to minimizing reconstruction error. While classical compressed sensing focuses on minimizing \( \| x - \hat{x} \|_2^2 \) (the reconstruction error), recent studies have demonstrated that mutual information offers a more general criterion for signal preservation, particularly in task-agnostic settings \cite{Zhang2018CSMI,Thiemann2016ITCS}. In these contexts, mutual information optimizes the relationship between the compressed and original signals to ensure that the compression process retains as much relevant information as possible, even when reconstruction is not the primary objective. 

We adopt the mutual information objective to ensure that the compressed representation retains semantically useful features, even without full pixel-wise recovery, which is more suitable for tasks where exact reconstruction is not always necessary \cite{Chen2019TaskAwareCS}.

\subsection{High-Efficiency Compressed Sensing Bottleneck Design}
To further optimize the compressed representation under limited supervision, we introduce a learnable parameterized distribution \( p_\theta(x|y) \), which models the probability of observing \( x \) given \( y \). This distribution helps approximate the true posterior distribution of the compressed signal. Specifically, we adopt a variational approach where the model learns a lower bound on the mutual information, guiding the encoder to preserve the most relevant information for downstream tasks. 

This transition from maximizing mutual information to using a likelihood-based approach is motivated by the intractability of computing mutual information directly in high-dimensional settings. Therefore, we use a variational lower bound to approximate the true mutual information, as proposed in \cite{Kingma2013VAE} and \cite{Berthelot2019MixMatch}.

The optimization objective of the High-Efficiency Compressed Sensing Bottleneck framework can be written as:
\begin{equation}
    \max_{\theta, \phi} \mathbb{E}_{Q_\phi(X, Y)} \big[\log p_\theta(x | y)\big].
\end{equation}
Since \( Q_{\text{data}}(X) \) not explicitly known, it is approximated using \( D \). Consequently, their gradients respect \( Q_{\text{data}}(X) \) which is computed by Monte-Carlo-sampling. The transforms the objective into a dataset-dependent formulation:
\begin{equation}
    \max_{\theta, \phi} \sum_{x \in D} \mathbb{E}_{Q_\phi(Y | x)} \big[\log p_\theta(x | y)\big] := \mathcal{L}(\phi, \theta; D).
\end{equation}

The feasibility is influenced by \( Q_\phi(Y | X) \). For instance, when \( Q_\phi(Y | X) \) is assumed to follow an isotropic Gaussian distribution \( \mathcal{N}(W f_\psi(X), \sigma^2 I_m) \), sampling is straightforward due to the well-defined properties of Gaussian noise.

For the parameter \( \theta \), Monte Carlo gradient estimates are efficiently calculated by leveraging the linearity of expectation. However, optimizing \( \phi \) poses additional difficulties, as it governs \( Q_\phi(Y | X) \) which is the sampling distribution. Tackle these challenges, we use with control variates to score the function gradient estimators\cite{fu2006gradient,glynn1990likelihood,williams1992simple} can be utilized. Reparameterization techniques are applicable to many continuous distributions, such as isotropic Gaussian and Laplace distributions. These methods involve transforming samples from a fixed base distribution via a deterministic function, enabling gradient estimates with reduced variance\cite{rezende2014stochastic,glasserman2013monte,schulman2015gradient,kingma2014autoencoding}.

We subsequently need to insert the HECSA into the DNN model by knowledge distillation method to form the bottleneck layer, and in turn design the entire HECS-B architecture.

\subsection{HECS-B Architecture Desgn}
To ensure the end-to-end trainability of our split computing model, we introduce multiple objectives that work at different parts of the pipeline. The first objective maximizes mutual information, ensuring that the compressed representation retains critical information. However, compression is only one aspect of our framework. To guide the decoder and enhance the quality of reconstructed signals, we will introduce knowledge distillation loss and its evolution later.

\subsubsection{Knowledge Distillation}

Model compression~\cite{hinton2014distilling} aimed at transferring knowledge from pre-trained ``teacher" model to ``student" model. Make student model could approximate the teacher’s performance while requiring less computational effort.

Given an input \( x \), \( T(x) \),\( S(x) \) denote the output logits from the large and small models. These logits are modulated by a temperature parameter \( \tau > 0 \):
\begin{equation}
    P_T(x) = \text{softmax}\left(\frac{T(x)}{\tau}\right),\quad
    P_S(x) = \text{softmax}\left(\frac{S(x)}{\tau}\right).
\end{equation}
The temperature \( \tau \) adjusts the smoothness of the probability distributions, where higher values produce softer probabilities that highlight inter-class relationships.

Training the student model involves a combined loss function that incorporates the cross-entropy loss for the true labels \( y \) and a distillation loss aligning the student’s distribution \( P_S(x) \) with the teacher’s distribution \( P_T(x) \). The total loss is expressed as:
\begin{equation}
    \mathcal{L}_{\text{total}} = (1 - \alpha) \mathcal{L}_{\text{CE}}(P_S(x), y) + \alpha \tau^2 \mathcal{L}_{\text{KL}}(P_T(x), P_S(x)).
\end{equation}
where \( \alpha \in [0, 1] \) balances the contributions of the two losses. Here, \( \mathcal{L}_{\text{CE}} \) represents crossentropy loss, \( \mathcal{L}_{\text{KL}} \) denotes the KullbackLeibler divergence:
\begin{equation}
    \mathcal{L}_{\text{KL}}(P_T, P_S) = \sum_{x} P_T(x) \log\left(\frac{P_T(x)}{P_S(x)}\right).
\end{equation}

This training process allows student model learning the output of the teacher model while also leveraging ground-truth labels for guidance. Knowledge distillation has gained popularity as a technique for enhancing the performance of compact models, particularly in resource-constrained environments.

\subsubsection{Integrating the Bottleneck into Model Splitting via Knowledge Distillation}

To incorporate the High-Efficiency Compressed Sensing Bottleneck into a segmentation model for distributed inference, a knowledge distillation framework is employed. The bottleneck acts as encoder which is \( q_\theta(z | x) \), the decoder which is \( p_\phi(h | z) \), both parameterized by neural networks, while introducing a trainable prior \( p_\phi(z) \) within the latent space. This design enables the bottleneck to generate compact and informative representations \( z \), which support distributed model inference.

The distillation objective aims to maximize mutual information \( z \), also the supervised target \( h \), while suppressing the influence of irrelevant features from the input \( x \). Given \( (x, h) \) which is a training pair, produced from original larger model, the objective function:
\begin{equation}
    \mathcal{L}(x, h) = - \mathbb{E}_{q_\theta(z | x)} \big[ \log p_\phi(h | z) - \beta \log p_\phi(z) \big],
\end{equation}
where the first term represents the reconstruction loss (distortion), and the second term, weighted by \( \beta \), controls the compression rate.

To facilitate efficient optimization, \( p(h | z) \) is modeled as a Gaussian distribution with mean given by a deterministic prediction \( g_\phi(z) \). Also, \( q_\theta(z | x) \) is assumed to follow a uniform distribution centered on the encoder output \( f_\theta(x) \), expressed by \( q_\theta(z | x) = \mathcal{U}(f_\theta(x) - \frac{1}{2}, f_\theta(x) + \frac{1}{2}) \). By leveraging the reparameterization trick~\cite{kingma2014autoencoding}, the loss function is reformulated by:
\begin{equation}
    \mathcal{L}(x, h) = \frac{1}{2} \| h - g_\phi(f_\theta(x) + \epsilon) \|_2^2 - \beta \log p_\phi(f_\theta(x) + \epsilon)
\end{equation}
where \( \epsilon \) accounts for the rounding operation during training, promoting robustness in optimization.

After training, the latent representation \( z \) is discretized as \( z = \lfloor f_\theta(x) \rceil \), enabling efficient entropy coding based on the prior \( p_\phi(z) \). The trainable prior \( p_\phi(z) \), inspired by neural image compression techniques~\cite{balle2018variational}, is factorized over the dimensions of \( z \), supporting scalable and parallel entropy coding. This integration ensures that the bottleneck delivers a compact and reliable representation, facilitating model splitting and distributed inference.

\section{Experiments and Evaluation}

\subsection{Performance Evaluation}

To assess the performance of the High-Efficiency Compressed Sensing Bottleneck (HECS-B), its results were compared with several baseline approaches across three datasets: MNIST~\cite{lecun1998gradient} and Omniglot~\cite{lake2015human}. the Omniglot dataset and the MNIST dataset images both have a consistent resolution of \( 28 \times 28 \). The following methods were evaluated:

\paragraph{LASSO with Random Gaussian Matrices.}  
This baseline~\cite{matsubara2022towards} utilized LASSO decoding as the measurement operator. With MNIST dataset, Omniglot dataset, which exhibit sparsity, additional transforms such as the DCT (which is Discrete Cosine Transform) provided no significant improvements.

\paragraph{Variational Autoencoder.}  
The Variational Autoencoder (VAE)~\cite{bora2017compressed} this approach which is a variable generative model. A mapping \( G: \mathbb{R}^k \to \mathbb{R}^n \) was defined, where \( G \) represents the observation model’s mean function. Reconstruction \( \hat{x} \) was computed by:
\[
    \hat{x} = G\left(\arg \min_z \|y - W G(z)\|_2\right),
\]
where the latent vector \( z \) was optimized to match the measurements \( y \) under \( G \). The architecture and parameters followed the defaults in~\cite{bora2017compressed}.

\paragraph{Implementation Details.}  
For HECS-B, the encoder was constrained to a linear transformation for direct comprise, which by random Gaussian matrices. And the HECS-Band VAE were implemented by perceptrons with two hidden layers, which multi layers, each containing 600 units. To ensure the robustness of \( W \) against unseen test signals, \( \ell_2 \)-regularization was applied to its norm, leading to a Lagrangian optimization formulation~\cite{grover2019uncertainty} which is:
\[
    \max_{\theta, \phi} \mathbb{E}_{Q_\phi(X, Y)} \big[\log p_\theta(x | y)\big], \quad \text{subject to } \|W\|_F \leq k.
\]
Here~\cite{grover2019uncertainty}, using line search to adjust Lagrangian parameter, with \( k \) set to Frobenius. This regularization ensured that HECS-Bdid not trivially increase \( W \) to mitigate noise effects.

Notably, the learned measurement matrix \( W \) in HECS-Bexhibited a significantly smaller norm than random Gaussian matrices, emphasizing that its performance improvements stemmed from the model's robustness and effectiveness rather than trivial scaling.

\begin{figure}
  \centering
  \begin{tikzpicture}
    \begin{axis}[
      width=0.8\textwidth,
      height=0.6\textwidth,
      xlabel={Number of Measurements (\( m \))},
      xlabel style={font=\small},
      ylabel={Reconstruction Error (\( \ell_2 \) norm)},
      ylabel style={font=\small},
      legend style={
        at={(0.5,-0.35)},
        anchor=north,
        legend columns=-1,
        font=\small
      },
      grid=both,
      xmin=0, xmax=105,
      ymin=0, ymax=0.5,
      xtick={2,5,10,25,50,100},
      ytick={0.05,0.1,0.2,0.3,0.4,0.5},
      tick label style={font=\small},
      legend cell align={left}
    ]

      \addplot[color=blue, mark=o, dashed, thick] coordinates {
        (2, 0.45) (5, 0.35) (10, 0.28) (25, 0.20) (50, 0.15) (100, 0.12)
      };
      \addlegendentry{LASSO}

      \addplot[color=red, mark=square, dash dot, thick] coordinates {
        (2, 0.40) (5, 0.30) (10, 0.24) (25, 0.18) (50, 0.13) (100, 0.10)
      };
      \addlegendentry{VAE}

      \addplot[color=green, mark=triangle*, thick] coordinates {
        (2, 0.25) (5, 0.18) (10, 0.12) (25, 0.08) (50, 0.05) (100, 0.03)
      };
      \addlegendentry{HECS-B}

    \end{axis}
  \end{tikzpicture}
  \caption{Reconstruction Errors across methods: comparison of LASSO, VAE, and HECS-B.}
  \label{fig:reconstruction-errors}
\end{figure}
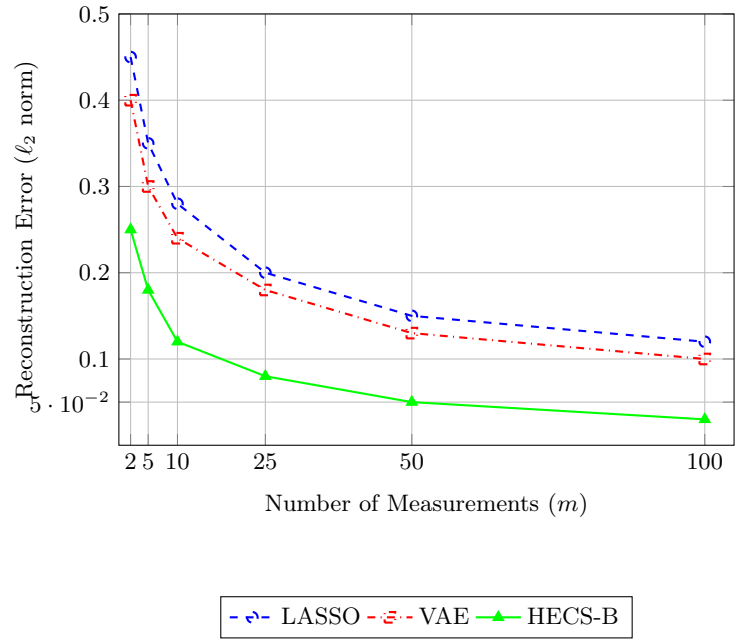
Figure~\ref{fig:reconstruction-errors} illustrates the \( \ell_2 \) reconstruction errors on MNIST and Omniglot test sets. Across all evaluated \( m \), HECS-Bconsistently outperformed both LASSO and VAE methods. The LASSO method (blue curves) struggled to reconstruct signals effectively with limited measurements, leading to high reconstruction errors. While the VAE method (red curves) achieved lower errors compared to LASSO, its improvement rate slowed as \( m \) increased. In contrast, HECS-B(green curves) demonstrated the best performance, achieving the lowest reconstruction errors across all tested configurations, thereby showcasing its capacity to preserve and recover critical data even under compression. For \( m = 25 \), HECS-Bachieved reconstructions closely resembling the original signals, highlighting its precision and robustness, which were unmatched by the baseline methods.

\subsection{Deployment Evaluation}
\subsubsection{Experimental Environment}

The deployment environment consists of a distributed collaborative inference framework, incorporating various edge devices and a central Jetson AGX Orin server. The detailed hardware specifications are in Table~\ref{tab:t1}
\begin{table}[t]
\caption{Simplified Device Specifications}
\label{tab:t1}
\resizebox{1\textwidth}{!}{
\centering
\begin{tabular}{|l|l|l|}
\hline
\textbf{Device} & \textbf{Compute Power} & \textbf{Category} \\
\hline
NVIDIA Jetson Nano & 128-core Maxwell GPU & Edge Device \\
NVIDIA Jetson Xavier NX & 128-core Volta GPU with 36 Tensor Cores (FP16 precision) & Edge Device \\
Raspberry Pi 4 Model B & Quad-core ARM Cortex-A72 CPU & Edge Device \\
NVIDIA Jetson AGX Orin & 2048-core Ampere GPU with 64 Tensor Cores, up to 275 TOPS & Central Host \\
\hline
\end{tabular}
}
\end{table}

Within this architecture, the Jetson AGX Orin operates as the primary inference server, utilizing its high computational capacity to manage complex and resource-heavy operations. The edge devices—comprising Jetson Nano, Jetson Xavier NX, and Raspberry Pi—are designated for initial data preprocessing, feature extraction, and partial inference.

The devices communicate use Secure Shell Protocol (SSH) through WIFI, ensuring low-bandwidths and low-data transmission speed. This setup replicates real-world deployment scenarios where edge devices with limited resources depend on a central server for enhanced processing capabilities.

To improve deployment efficiency, the HECS-B is implemented on the edge devices and server. This method enables effective feature compression, reduces bandwidth demands for data transmission, and ensures high accuracy during inference.

\subsubsection{Experimental Setup}

To assess the performance of the HECS-B framework, experiments were conducted on a large-scale image classification task within a distributed collaborative inference environment. The ImageNet (ILSVRC 2017) dataset~\cite{russakovsky2015imagenet}, comprising 1.29 million training images and 60,000 validation images, was employed for this evaluation. Following standard experimental protocols, the models were trained on the ImageNet dataset, use top-1 to evaluate classification accuracy which was computed on the validation set.

\subsubsection{Model Configuration} The backbone network for this study was ResNet-50~\cite{he2016deep}, which is a model pre-trained on the ImageNet dataset. To incorporate HECS-B framework, layers preceding the third residual block were substituted by HECS-B modules utilizing neural compression techniques. These modules compress intermediate features at the encoder stage and reconstruct them at the decoder for subsequent processing. During the initial training phase, the encoder-decoder modules were trained to mimic output of corresponding residual block form original ResNet-50 (original large model). Second phase, the entire model, including the encoder, decoder, and all remaining layers, was fine-tuned to optimize classification performance.

The model training followed a two-stage approach:
\begin{itemize}
    \item Teacher-Student Training: Knowledge distillation was employed to train the encoder-decoder modules, aligning their outputs \( h \) with those of original large model residual block outputs.
    \item Fine-Tuning: For submodel network, within HECS-B module, was subsequently fine-tuned end-to-end to enhance performance on the classification task.
\end{itemize}

\subsubsection{Baseline Methods for Comparison} To benchmark HECS-B's effectiveness, its performance was compared against several compression baselines:
\begin{itemize}
    \item JPEG and WebP: Popular image compression formats widely used in real-world applications~\cite{matsubara2022towards}.
    \item CR+BQ: which Combines the channel reducing, and bottleneck quantize~\cite{matsubara2022towards}.
    \item Neural Compression: Includes factorized prior~\cite{balle2018variational}, mean-scale hyperprior~\cite{minnen2018joint}, commonly utilized for neural image compression tasks.
    \item Shallow Variational Bottleneck (SVB): Employs a simplified neural network architecture for variational compression~\cite{furtuanpey2024frankensplit}. Unlike HECS-B, it applies a single variational layer for feature compression, resulting in reduced model complexity but lower accuracy.
\end{itemize}

The input to the ResNet-50 backbone was formatted as a tensor of dimensions \( 3 \times 224 \times 224 \), following standard preprocessing protocols for ImageNet. Rate-distortion (RD) performance was analyzed using supervised RD curves, with x axis denoting mean data size, y axis indicating top-1 accuracy.

\begin{figure}[htbp]
    \centering
    \includegraphics[width=1\textwidth]{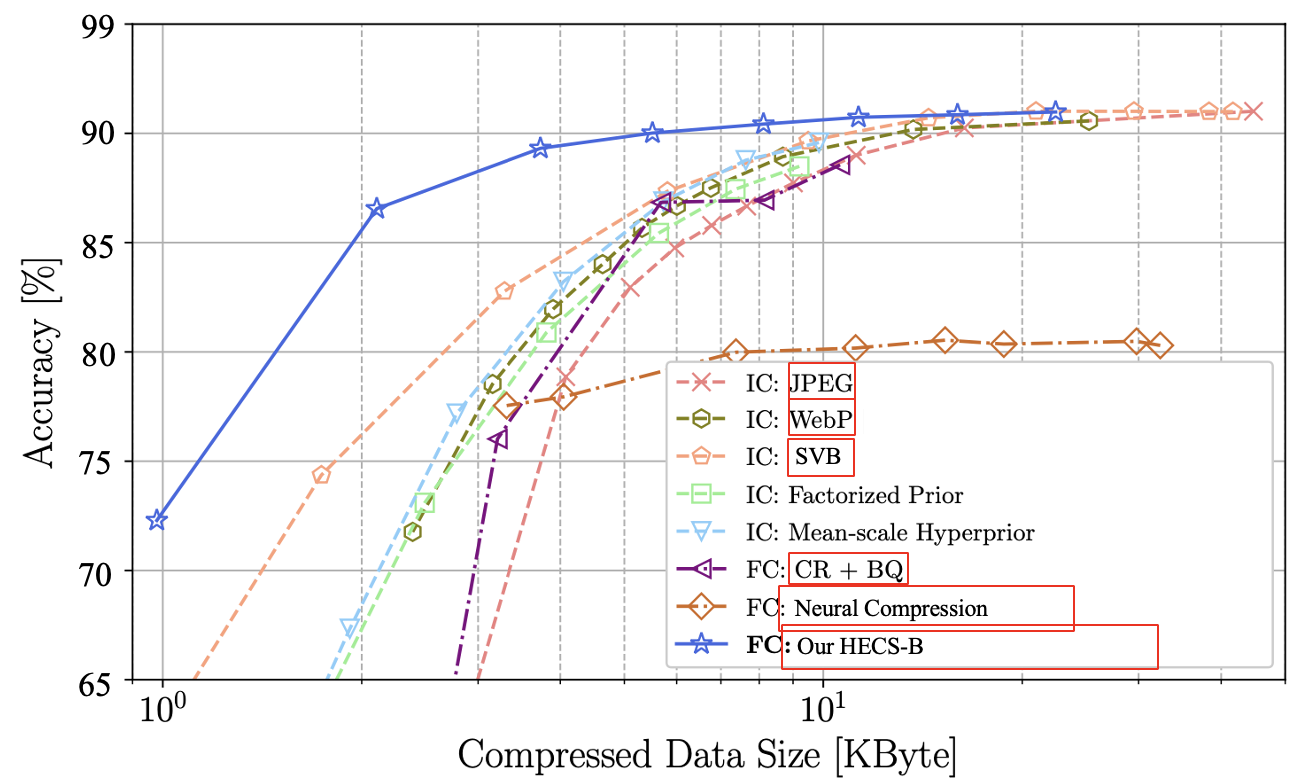}  
    \caption{Rate-distortion (RD) performance curves.}
    \label{fig:f9}
\end{figure}

\begin{table}
  \centering
  \caption{Transfer and inference time under different data rates and codecs.}
  \label{tab:t2}
  \resizebox{1\textwidth}{!}{
  \begin{tabular}{llccc}
    \toprule
    \textbf{Network/Data Rate} & \textbf{Codec} & \textbf{Transfer (ms)} & \textbf{Total [Nano] (ms)} & \textbf{Total [NX] (ms)} \\
    \midrule
    \multirow{6}{*}{4G / 12.0 Mbps}
      & SVB & 22.21 & 41.09 & 40.15 \\
      & CR + BQ & 24.72 & 44.61 & 44.66 \\
      & Neural Compression & 36.85 & 52.89 & 52.01 \\
      & WebP & 40.48 & 62.63 & 62.63 \\
      & PNG & 56.98 & 70.13 & 70.13 \\
      & \textbf{HECS-B} & \textbf{10.90} & \textbf{20.20} & \textbf{20.20} \\
    \midrule
    \multirow{6}{*}{Wi-Fi / 54.0 Mbps}
      & SVB & 4.71 & 22.60 & 21.65 \\
      & CR + BQ & 5.05 & 24.93 & 23.99 \\
      & Neural Compression & 8.74 & 28.78 & 28.70 \\
      & WebP & 10.33 & 30.47 & 30.47 \\
      & PNG & 12.66 & 33.81 & 33.81 \\
      & \textbf{HECS-B} & \textbf{2.20} & \textbf{11.92} & \textbf{11.01} \\
    \midrule
    \multirow{6}{*}{5G / 66.9 Mbps}
      & SVB & 3.58 & 20.46 & 20.01 \\
      & CR + BQ & 4.85 & 22.73 & 21.56 \\
      & Neural Compression & 7.41 & 27.44 & 25.56 \\
      & WebP & 9.09 & 29.10 & 29.10 \\
      & PNG & 10.22 & 33.36 & 33.36 \\
      & \textbf{HECS-B} & \textbf{1.86} & \textbf{11.85} & \textbf{10.31} \\
    \bottomrule
  \end{tabular}
  }
\end{table}
\subsubsection{Experimental results} Figure~\ref{fig:f9} shows that HECS-B delivers exceptional performance at terms for rate-distortion efficiency. HECS-B outperforms baseline methods like JPEG, WebP, CR+BQ, Neural Compression and SVB by achieving better feature compression and maintaining high accuracy. While SVB offers a simpler alternative with lower computational demands, it is surpassed by HECS-B in both compression effectiveness and accuracy, highlighting the benefits of the advanced compressed sensing methodology.

We also test the transfer and inference time with different baseline method and our HECS-B method in different data rate. The results has show in Table~\ref{tab:t2}, and compare with all the baseline, HECS-B has show the outperforms, specially, with compare with SVB, our architecture HECS-B reduces bandwidth utilization by {50\%}, maintains high accuracy, and achieves a {60\% speed-up} in computational efficiency.

\section{Conclusions}

This work is recognized as a significant advancement in the field of split computing by addressing two critical challenges: bandwidth efficiency and real-time performance. The introduction of the High-Efficiency Compressed Sensing Bottleneck (HECS-B), inspired by compressed sensing theory, has redefined the approach to intermediate feature transmission in SC. A groundbreaking 50\% reduction in bandwidth utilization, along with a 60\% improvement in computational efficiency, has been achieved without compromising model accuracy. These results decisively surpass state-of-the-art SC methods, demonstrating the unparalleled effectiveness and scalability of the proposed framework.

The reliability and practicality of HECS-B have been firmly established through rigorous theoretical analysis and comprehensive experimental validation in both simulated and real-world environments. The proposed architecture is shown to bridge the gap between resource-constrained edge devices and computationally intensive cloud services, offering a robust solution for deploying advanced AI applications across diverse real-world scenarios. Key limitations of existing SC methods are addressed, and a foundation for future innovations in scalable and efficient edge-cloud computing is laid by this work.

%
%
%
%
\bibliographystyle{splncs04}
\bibliography{reference}

\end{document}